\documentclass{article}

\usepackage[final]{neurips_2020_ml4ps}

\usepackage[utf8]{inputenc} % allow utf-8 input
\usepackage[T1]{fontenc}    % use 8-bit T1 fonts
\usepackage{hyperref}       % hyperlinks
\usepackage{url}            % simple URL typesetting
\usepackage{booktabs}       % professional-quality tables
\usepackage{amsfonts}       % blackboard math symbols
\usepackage{nicefrac}       % compact symbols for 1/2, etc.
\usepackage{microtype}      % microtypography
\usepackage{amsmath,amsfonts,bm}
\usepackage{xspace}
\usepackage{cleveref}
\usepackage{natbib}%[round]
\usepackage{graphicx}
\usepackage{xcolor}
\usepackage{subcaption}

\newcommand{\C}{\mathbb{C}}

\newcommand{\ie}{{\it i.e.}\xspace}

\renewcommand{\epsilon}{\varepsilon}

\newcommand{\rhoss}{\rho^{ss}}
\newcommand{\vtheta}{\bm{\theta}}
\newcommand{\cL}{{\cal L}}
\newcommand{\tauz}{\langle \sigma_{z} \rangle}
\title{Inverse design of dissipative quantum steady-states with implicit differentiation}

\author{%
Rodrigo A.~ Vargas-Hern\'{a}ndez \thanks{University of Toronto. $^\dagger$Chemical Physics Theory Group, Department of Chemistry. $^\ddagger$Vector Institute.} 
\footnotemark[2] \footnotemark[3] \\
\texttt{r.vargashernandez@utoronto.ca} \\
\And
Ricky T. Q. Chen \footnotemark[1] \footnotemark[3] \\
\texttt{rtqichen@cs.toronto.edu}
\AND
Kenneth A. Jung \footnotemark[1] \footnotemark[2] \\
\texttt{kenneth.jung@utoronto.ca}
\And 
Paul Brumer \footnotemark[1] \footnotemark[2] \\
\texttt{paul.brumer@utoronto.ca}
}

\begin{document}

\maketitle

\begin{abstract}
Inverse design of a property that depends on the steady-state of an open quantum system is commonly done by grid-search type of methods.
In this paper we present a new methodology that allows us to compute the gradient of the steady-state of an open quantum system with respect to any parameter of the Hamiltonian using the implicit differentiation theorem.
As an example, we present a simulation of a spin-boson model where the steady-state solution is obtained using Redfield theory.
\end{abstract}

% ---------------------------
\section{Introduction}
The field of open quantum systems (OQS) is focused on studying the interaction of a \emph{system} (S) with its surroundings, typically referred to as the \emph{bath} (B) \citep{OQS_book}. The Hamiltonian for the complete problem is defined as the sum of the isolated system Hamiltonian, $H_{S}$, the bath that surrounds the system, $H_{B}$, and the interaction between the system and the bath, $H_{SB}$,
\begin{equation}
    H = H_S + H_B + H_{SB}.
\end{equation}
For physical problems the bath is often represented as a continuum that describes the vibrational or solvent degrees of freedom.
Given the large dimensionality of the composite Hilbert space of such systems, it is common to construct a reduced representation by tracing out the bath degrees of freedom, $\rho_{S} = \text{Tr}_{B}[\rho]$.
Here, we refer to $\rho_{S}$ as the reduced density matrix (RDM) and $\rho$ is the density matrix of the complete system.

In OQS, one is usually interested in the dissipative effects the bath induces in the system over time. %$\partial \rho_{S} (t)/\partial t = \text{Tr}_{B} \left \{\partial \rho /\partial t  \right \}$.
However, for many OQS the properties of interest are related to the steady-state (SS) of the system ($\rhoss$), $\; d \rho(t) / d t = 0$;
for example, energy transfer efficiency in biological systems excited by natural incoherent light \citep{Timur_PRL,Timur_JCP,Pachon_NJP} and the performance of quantum heat engines/refrigerators \citep{Dvira_PRE,QRef_PRL}.
Here, we propose a novel numerical algorithm to compute the gradient of the steady-state of an OQS with respect to any parameter of the Hamiltonian ($\vtheta_i$) using implicit differentiation,  $\partial \rho^{ss}/\partial \vtheta_i$, allowing us to tackle problems related to inverse design and sensitivity analysis. 

In the following sections, we discuss how as an example, we differentiate the steady-state of the spin-boson (SB) model with respect to any parameter using an ordinary differential equation solver and implicit differentiation.

\section{Differentiation of the steady-state}\label{sec:ss_diff}
Conditioned on initial values $\rho(t_0) \in \C^d$ and free parameters $\vtheta$, let $\rho(t) \in \C^d$ be the solution to a homogeneous dynamical system parameterized by an ordinary differential equation (ODE) $\frac{d\rho}{dt} = f(\rho, \vtheta)$ where $d$ is the dimensionality of the RDM. In optimization, we want to compute the gradient of the steady state with respect to the parameters $\vtheta$. Let us denote $\rhoss$ as the steady-state, which satisfies $f(\rhoss, \vtheta) = 0$.

We can solve for $\rhoss$ by running an appropriate ODE solver for a sufficiently long period of time. However, differentiating through the internals of the ODE solver is computationally expensive as it requires storing all intermediate quantities of the solver. The adjoint method for computing gradients of ODE solutions requires either the trajectory $\rho(t)$ to be stored in memory or solving $\rho(t)$ in reverse time, which uses constant memory~\citep{chen2018neural}. However, since we are interested in $\rhoss$, the reversing approach is not applicable as the steady state, once reached, cannot be reversed. Instead, since $\rhoss$ is the solution of a fixed point problem, the Jacobian $\frac{d\rhoss}{d\vtheta}$ can be expressed using the implicit function theorem~\citep{krantz2012implicit},
\begin{equation}\label{eq:implicit_fn_thm}
    \frac{d\rhoss}{d\vtheta} = - \left( \frac{df(\rhoss, \vtheta)}{d\rho} \right)^{-1} \left[ \frac{d f(\rhoss, \vtheta)}{d\vtheta} \right].
\end{equation}
We show that it is possible to compute this gradient, i.e. Eq. (\ref{eq:implicit_fn_thm}), using constant memory cost, and without the need to know how $\rhoss$ is computed as long as it satisfies the steady state criterion, \ie $f(\rhoss, \vtheta) = 0$.

%The optimization of hyperparameters of machine learning (ML) models through implicit differentiation can be done as was demonstrated by \citep{ML_implicit_diff}.
%In the context of optimization, 
For this work, we would also have a scalar loss function $ \cL(\vtheta,\rhoss)$ that we wish to minimize. Here, the gradient with respect to parameters can be factored with the chain rule $\frac{d\cL}{d\rhoss} \frac{d\rhoss}{d\vtheta}$. In general, we're interested in vector-Jacobian products of the form $v \frac{d\rhoss}{d\vtheta}$ where $v$ is any vector. 
In automatic differentiation libraries such as jax~\citep{jax2018github}, we can easily compute vector-Jacobian products for large systems but recovering the full Jacobian is more costly.

Computing a vector-Jacobian product using, Eq. (\ref{eq:implicit_fn_thm}), requires solving $vJ^{-1}$ where $J = \left( \frac{df(\rhoss, \vtheta)}{d\rho} \right)^{-1}$. For large systems where $J$ cannot be tractably computed, we compute $vJ^{-1}$ by recognizing that it is the steady-state solution of the ODE $\frac{dy}{dt} = yJ - v$. Notably, simulating this ODE only requires vector-Jacobian products, which are inexpensive in automatic differentiation libraries. 
%Thus, in order to compute the gradient of a steady-state system, we solve for the another steady-state system.
In the following sections we illustrate how inverse design and sensitivity analysis are possible for dissipative quantum systems by differentiating through the steady state of a Spin-Boson model. 

% Add related works to implicit MAML, Jon Loraine's work, optnet, deep equilibrium models, etc.
% Compared to these works that differentiate through the solution of an optimization problem, the steady state we are interested in depends on the initial conditions. As such, we make use an ODE solver to compute the steady state, and since we have it handy, use it compute the implicit fn theorem.

\section{Example: Spin-Boson model}
In the field of OQS, the SB Hamiltonian \citep{Thoss2001} is a standard model used to describe a wide variety of physical phenomena \citep{OQS_book,RevModPhys_1987} including electron transfer \citep{Redfield_theory}, heat transport \citep{Xu2016}, and energy transfer \citep{Liu2019,Jung2020}.
The total Hamiltonian is,
\begin{eqnarray}
    H_S = \frac{\epsilon}{2}\sigma_z + \frac{\Delta}{2}\sigma_x, \;\;\;\; H_B = \sum_k \omega_k b_k^\dagger b_k,  \;\;\;\; H_{SB} = \sigma_z \sum_k \lambda_k (b_k^\dagger + b_k),
\end{eqnarray}
where $b_k^\dagger$ ($b_k$) is the creation (annihilation) operator of mode $k$ in the bath, $\sigma_{z}$ and $\sigma_x$ are Pauli matrices, and $\{\lambda_k\}$ are the coupling strength parameters, and $\epsilon$ and $\Delta$ are system parameters.

%In the Redfield theory (RT) \citep{Redfield,Redfield_theory}, the equations of motions of the reduced density matrix elements are,
The exact equation of motion \citep{Nakajima1958,Zwanzig1960} for the RDM is not practical to solve since one requires knowledge of the interacting system-bath dynamics.
Redfield theory (RT) \citep{Redfield,Palenberg2001} is a practical alternative for regimes in which the system-bath coupling is weak, i.e. $\{\lambda_k\}\ll1$. The RT equations of motions written in the system eigenbasis are,
% \begin{eqnarray}
%     \frac{\partial \rho_{S}(t)}{\partial t} = -i\left [ H_S,\rho_{S}(t)) \right ] - \int_0^t \mathrm{d} \tau \; \text{Tr}_{B}\left \{ \left [ H_{SB},\left [ H_{SB}(\tau),\rho_{S}(t)\otimes \rho_{B}  \right ] \right ] \right \},
% \end{eqnarray}
% where $[,]$ is the commutator operator, and $\rho_{B}$ is the density matrix that describes the bath. $H_{SB}(\tau)$ is the interaction picture representation of the system-bath interaction Hamiltonian. 
% For the SB model, $\rho_{B}$ is the equilibrium density matrix, $\rho_{B} = \exp^{-H_{B}\beta}/\text{Tr}_{B}\left\{\exp^{-H_{B}\beta} \right\}$.
%\begin{eqnarray}
%    \rho_{B} = \frac{\exp^{-H_{B}\beta}}{\text{Tr}_{B}\left\{\exp^{-H_{B}\beta} \right\}}.
%\end{eqnarray}
% In the RT formalism, the equations of motion of $\rho_{S}(t)$ are commonly described in terms of the eigenstates of the system, $H_S |\mu\ \rangle = \epsilon_{\mu} |\mu \rangle $.
% The equations of motions are, 
\begin{eqnarray}
    \frac{\partial \rho_{\mu,\nu}(t)}{\partial t} = -i\omega_{\mu,\nu}\rho_{\mu,\nu}(t) + \sum_{\kappa,\lambda}R_{\mu,\nu,\kappa,\lambda}\;\rho_{\kappa,\lambda}(t),
\end{eqnarray}
where $[\mu,\nu,\kappa,\lambda]$ are the index of the eigenstates of $H_S$, i.e. they satisfy $H_S |\mu\ \rangle = \epsilon_{\mu} |\mu \rangle $, and $\omega_{\mu,\nu}$ is the difference between eigenvalues $\epsilon_\mu$ and $\epsilon_\nu$.
$R_{\mu,\nu,\kappa,\lambda}$ are the Redfield tensors which describe the interaction of the system and bath and are given by, 
\begin{eqnarray}
    R_{\mu,\nu,\kappa,\lambda} = \Gamma^{+}_{\lambda,\nu,\mu,\kappa} + \Gamma^{-}_{\lambda,\nu,\mu,\kappa} - \delta_{\nu,\lambda}\sum_{\alpha}\Gamma^{+}_{\mu,\alpha,\alpha,\kappa} - \delta_{\mu,\kappa}\sum_{\alpha}\Gamma^{+}_{\lambda,\alpha,\alpha,\nu},
\end{eqnarray}
which contain the transition rates, 
{\normalsize
\begin{eqnarray}
\Gamma^{+}_{\lambda,\nu,\mu,\kappa} &=& \langle \lambda|\sigma_z |\nu \rangle \langle \mu|\sigma_z |\kappa \rangle \int_{0}^{\infty} \mathrm{d} \tau F(\tau)e^{-i\omega_{\mu,\kappa}\tau}, \\%\;\;\text{ and }\;\;
\Gamma^{-}_{\lambda,\nu,\mu,\kappa} &=& \langle \lambda|\sigma_z |\nu \rangle \langle \mu|\sigma_z |\kappa \rangle \int_{0}^{\infty} \mathrm{d} \tau F^*(\tau)e^{-i\omega_{\lambda,\nu}\tau},
\end{eqnarray}}
that are in turn comprised of the bath correlation function,
\begin{eqnarray}
F(\tau) = \int_{0}^{\infty} \mathrm{d} \omega \; g(\omega)\left [ \coth(\beta\omega/2)\cos(\omega\tau) - i\sin(\omega\tau) \right ].
\end{eqnarray}
$g(\omega)$ is the spectral density function, $g(\omega) = \eta \omega^s \omega_c^{1-s} e^{-\omega/\omega_c}$; for this work we used the super Ohmic form of $s=3$. 
$\eta$ is the bath friction parameter that is on the order of $\lambda_k^2$ and $\beta$ is the inverse temperature.
%$(F(\tau))^*$ is the complex-conjugate of $F(\tau)$. 

\section{Results}
\subsection{Sensitivity analysis}
\begin{figure}[h!]
    \centering
    \includegraphics[scale=0.33]{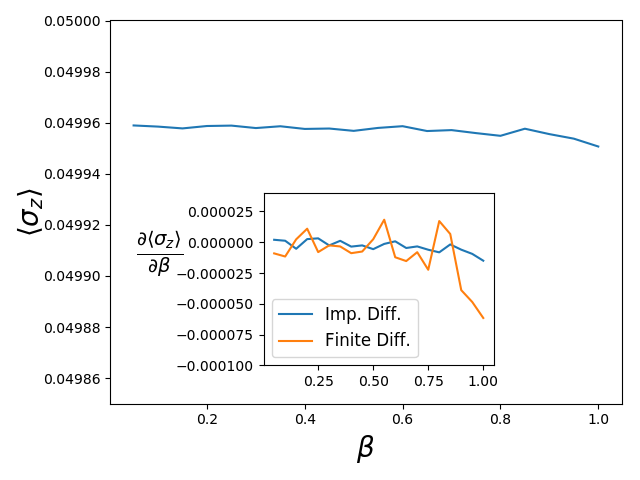}
    \includegraphics[scale=0.33]{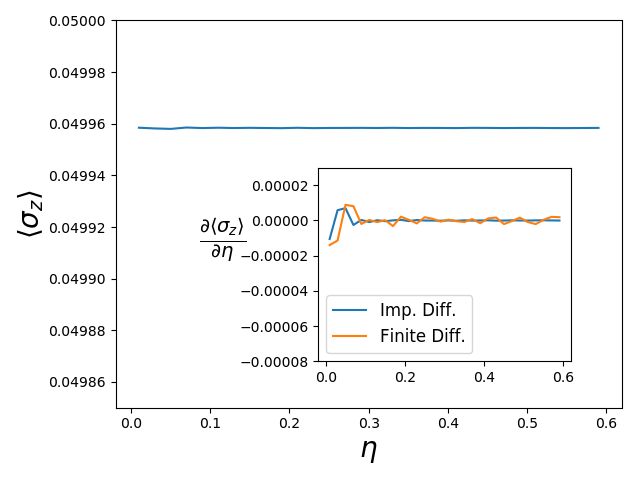}
    \includegraphics[scale=0.33]{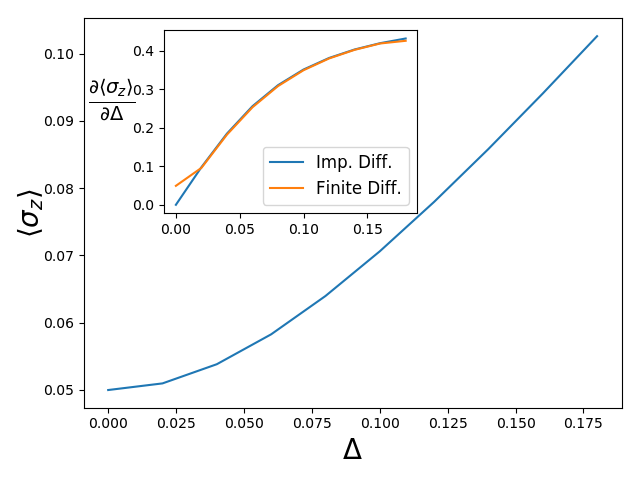}
    \caption{$\tauz =  \text{Tr}[\sigma_{z}\rhoss]$ as a function of different parameters, $\beta$, $\eta$ and $\Delta$. The inset of each figure compares the gradient computed with Eq. (\ref{eq:implicit_fn_thm}) (blue solid curve)  and finite differences (orange solid curve). 
    For all calculations, except for the parameter in play, we used $\beta = 0.1$, $\eta = 0.01$, $w_c = 1$, $\epsilon = 0.1$, and $\Delta = 0$. The initial density matrix used was $\rho_S(t_0) = [3/4, -i\frac{\sqrt{3}}{4},i\frac{\sqrt{3}}{4},1/4]$ (Eq. \ref{eqn:rho_t0}). }
    \label{fig:tau_z_1d}
\end{figure}

One of the most studied observables for the SB model is the population difference at equilibrium, $\tauz = \text{Tr}[\sigma_{z}\rhoss]$. 
Fig. (\ref{fig:tau_z_1d}) illustrates the change of $\tauz$ with respect to bath parameters $\beta$, $\eta$ and the system parameters $\Delta$. For all calculations we used $w_c = 1$, $\epsilon = 0.1$ and the initial density matrix was taken to be
\begin{eqnarray}
\rho_S(t_0) = \begin{pmatrix}
\frac{3}{4} & -i\frac{\sqrt{3}}{4}\\ 
 i\frac{\sqrt{3}}{4}& \frac{1}{4}
\end{pmatrix}.
\label{eqn:rho_t0}
\end{eqnarray}
%\rho_S(t_0) = [0.75, -i\frac{\sqrt{3}}{4},i\frac{\sqrt{3}}{4},0.25]$
We can observe that $\tauz$ is independent of $\beta$ and $\eta$, as the gradients are effectively zero.
However, $\frac{\partial \tauz}{\partial \Delta}$ is nonzero which confirms the effect $\Delta$ has in $\tauz$. 

While finite differences with respect to $\beta$, $\eta$, and $\Delta$ can be computed one at a time, this requires $\mathcal{O}(d)$ evaluations of the steady state.
The automatic differentiation process discussed in Section~\ref{sec:ss_diff} allows us to compute analytical derivatives for all parameters simultaneously with just one evaluation of the steady-state.

To construct the Redfield tensors we must know the eigenvalues and eigenstates of $H_S$ beforehand. 
%In the SB model, $H_{S}$ is not diagonal for $\Delta > 0$. 
As $H_{S}$ is not diagonal for $\Delta > 0$, the eigenvalues and eigenstates do not have trivial analytical solutions. 
To compute these, we used jax's eigendecomposition. %which also provided derivatives.
This ensures this method can be applied to arbitrary systems.
%We emphasize that no analytic solutions for the eigenvalues or eigenstates were used during the calculations. 
%This can be done in the jax library which ensures generality of the approach for arbitrary systems. 

\break
\subsection{Inverse design for system's Hamiltonian}

Gradient based methods have proven to be powerful numerical tools to find the minimizer of loss functions. 
Given the possibility to compute quantities like $\partial \cL/\partial \theta_i$, we could reformulate the inverse design problem into an optimization one. For example, what values of $\epsilon$ and $\Delta$ reproduce a target observable. In order to do so we define an error function, e.g., $\cL(\epsilon,\Delta)=\left \| \tauz - \Hat{\tauz}  \right \|_{2}$.

\begin{figure}[h!]
    \centering
    \begin{subfigure}[b]{0.5\linewidth}
    \centering
    \includegraphics[scale=0.45]{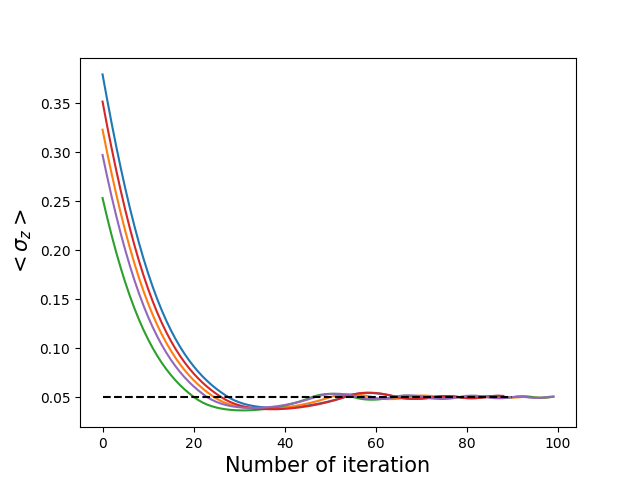}
    \caption{}
    \end{subfigure}%
    \begin{subfigure}[b]{0.5\linewidth}
    \centering
    \includegraphics[scale=0.45]{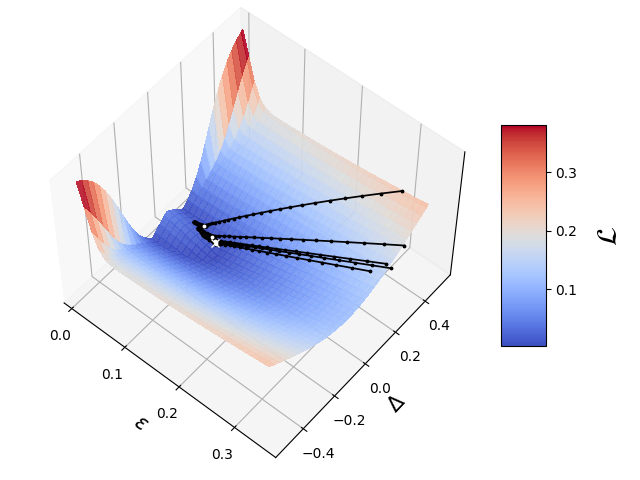}
    \caption{}
    \end{subfigure}
    \caption{We optimize the parameters of $H_S$, $\epsilon$ and $\Delta$, with respect to a target $\hat{\tauz}$ computed with $\epsilon = 0.1$ and $\Delta = 0$. We used the Adam optimization algorithm to minimize $\cL(\epsilon, \Delta)=\left \| \tauz - \Hat{\tauz}  \right \|_{2}$. \textbf{(a)} The value of $\tauz$ at each iteration of the optimization routine for different random initialization. The black-dashed curves is the target $\tauz$. \textbf{(b)} The iterations of the optimization procedure in the parameter space for different random initialization. The white-$x$ symbol represents the target set of parameters and the white-$\bullet$ is the set of parameters found at 100th iteration of the Adam algorithm. $\frac{\partial \cL}{\partial \epsilon}$ and $\frac{\partial \cL}{\partial \Delta}$ used in the Adam optimizer were computed with Eq. (\ref{eq:implicit_fn_thm}). For all calculations, we used $\beta = 0.1$, $\eta = 0.01$, $w_c = 1$ and $\rho(t_0)$ was the same as in Fig. (\ref{fig:tau_z_1d}). The learning rate was set to 0.1, and we parameterized $\epsilon$ and $\Delta$ to be strictly positive by using the softplus function.}
    \label{fig:tau_z_opt}
\end{figure}

Fig. \ref{fig:tau_z_opt} illustrates how gradient-based methods can be used to search for the optimal values of  $\epsilon$ and $\Delta$ that reproduce a target $\tauz$; as a proof of principle calculation we used $\epsilon = 0.1$, $\Delta = 0$ where $\tauz= 0.04995847$. 
For these simulations we fixed the values of $\beta$ and $\eta$. 
Furthermore, in Table \ref{tab:Adam} we report the values of $\epsilon$ and $\Delta$ for 5 different optimizations.
To minimize $\cL$ we used the Adam algorithm, which is a first-order gradient-based optimization algorithm \citep{Adam}.

\begin{table}[h!]
    \caption{$\;$}
    \centering
    \begin{tabular}{c c | c |c}
    $\epsilon$ & $\Delta$ & $\tauz$ & $\cL$ \\ \hline\hline
    0.0835 & 0.0578 & 0.049947 & 1.2E-5  \\
    0.0867 & 0.0526 & 0.049942  & 1.6E-5  \\
    0.0739 & 0.0692 & 0.049913 & 4.5E-5 \\
    0.0490 & 0.0809 & 0.049921 & 3.8E-5 \\
    0.0822 & 0.0597 & 0.049940 & 1.9E-5
    \end{tabular}
    \label{tab:Adam}
\end{table}

\section{Summary}
We have presented a novel numerical methodology capable of computing the gradient of quantum observables of the form $\langle \hat{O} \rangle = \text{Tr}[\hat{O} \rhoss]$, with respect to any parameter of the Hamiltonian.
As we stated, this procedure does not depend on the numerical procedure used to solve for $\rhoss$ and it uses constant memory.
By computing the gradient of $\tauz$, for the particular parameter set and model used we found an independence of $\tauz$ with respect to the bath parameters. 
$\tauz$ increases as a function of $\Delta$ and the gradient computed with Eq. (\ref{eq:implicit_fn_thm}) matches the gradient computed with finite differences. 

Inverse design for physical systems is one of the most common applications for gradient-based search algorithms.
We have demonstrated that the Adam algorithm can be used to find the optimal values of $\epsilon$ and $\Delta$ that reproduce a target $\tauz$.

% Future work.
For all the results presented here, we considered a unique $\rho_{S}(t_0)$; however, one can apply this technique to search for the optimal $\rho_{S}(t_0)$ given a scalar observable that depends on the steady-state.
For example, in the SB model one can define %$\vtheta = [\rho_{00}(t_0),\rho^{R}_{01}(t_0),\rho^{I}_{01}(t_0),\rho^{R}_{10}(t_0),\rho^{I}_{10}(t_0),\rho_{11}(t_0)]$; $\rho^{R}_{ij}$ and $\rho^{I}_{ij}$ are the real and imaginary parts of the off-diagonal elements of the density matrix.
$\vtheta$ to include the initial values of the RDM to study the dependence of the SS on initial conditions.

% \section{Things to do}
% \begin{itemize}
%     \item Gradients with respect to $\beta$, $\eta$ ($\epsilon = 0.1$, $\Delta=0.$, $\rho=[0.75,0.,-\sqrt{3}/4,0.,-\sqrt{3}/4,0.25]$)
%     \item Gradient descent optimization for $\langle \tau_z \rangle = \rho_{00} - \rho_{11}$
%     \item 
% \end{itemize}

\section{Broader Impact}
Open quantum systems appear in many areas in physics. Our work can potentially be used for designing quantum heat engines or materials with specific desirable properties. This work itself is unlikely to have immediate ethical issues.

% Will be included in the Final paper
\section*{Acknowledgement}
Support by the US Air Force (AFOSR) under  grant number FA9550-17-1-0310 is gratefully acknowledged.

{%\small
\bibliographystyle{plainnat}
\bibliography{references}
}

\end{document}